\documentclass[12pt]{article}
\usepackage[dvips]{graphicx}
\usepackage{amssymb}
\usepackage{amsmath}

\newcommand{\bea}{\begin{equation}}
\newcommand{\eea}{\end{equation}}
\newcommand{\be}{\begin{eqnarray}}
\newcommand{\ee}{\end{eqnarray}}
\newcommand{\nn}{\nonumber}
\newcommand{\ev}{\mbox{eV}}
\newcommand{\mev}{\mbox{MeV}}
\newcommand{\gev}{\mbox{GeV}}
\newcommand{\tev}{\mbox{TeV}}
\def\hbar#1{\backslash\hspace{-2mm}#1}

\def\nn{\nonumber}

\def\2tvec#1#2{
\left(
\begin{array}{c}
#1  \\
#2  \\
\end{array}
\right)}

\def\mat2#1#2#3#4{
\left(
\begin{array}{cc}
#1 & #2 \\
#3 & #4 \\
\end{array}
\right) }

\def\Mat3#1#2#3#4#5#6#7#8#9{
\left(
\begin{array}{ccc}
#1 & #2 & #3 \\
#4 & #5 & #6 \\
#7 & #8 & #9 \\
\end{array}
\right) }

\def\3tvec#1#2#3{
\left(
\begin{array}{c}
#1  \\
#2  \\
#3  \\
\end{array}
\right)}

\def\hbar#1{\backslash\hspace{-2mm}#1}

\numberwithin{equation}{section}

\begin{document}

\begin{titlepage}
\begin{center}

\vspace{1cm}
{\large\bf $T_{13}$ Flavor Symmetry and Decaying Dark Matter }
\vspace{1cm}

Yuji Kajiyama$^{a,b}$\footnote{yuji.kajiyama@kbfi.ee}
and
Hiroshi Okada$^{c,}$\footnote{HOkada@Bue.edu.eg}
\vspace{5mm}

{\it%
$^{a}${National Institute of Chemical Physics and Biophysics,\\[-1.5mm]
Ravala 10, Tallinn 10143, Estonia}\\
$^{b}${Department of Physics,~Niigata University,~Niigata 950-2128,~Japan}\\
$^{c}${Centre for Theoretical Physics, The British University in 
Egypt,\\[-1.5mm] El Sherouk City, Postal No, 11837, P.O. Box 43, Egypt}}

\vspace{8mm}

\abstract{We study a new flavor symmetric model with non-Abelian discrete symmetry $T_{13}$. 
The $T_{13}$ group is isomorphic to $Z_{13}\rtimes Z_3$, and it is the minimal group having two complex triplets in the irreducible representations. 
We show that the $T_{13}$ symmetry can derive lepton masses and mixings consistently. 
Moreover, if we assume a gauge-singlet fermionic decaying dark matter, its decay operators are also constrained by the $T_{13}$ symmetry so that only dimension six operators of leptonic decay are allowed. We find that the cosmic-ray anomalies reported by PAMELA and Fermi-LAT are well explained by decaying dark matter controlled by the $T_{13}$ flavor symmetry. }

\end{center}
\end{titlepage}

\setcounter{footnote}{0}

\section{Introduction}
Despite the great success of the Standard Model (SM) of the elementary particle physics, 
the origin of flavor structure, masses and mixings between generations, of matter particles are unknown yet. 
In order to overcome these problems, plenty of models based on the principle of symmetry, flavor symmetry, have been discussed. In particular, the fact that the lepton mixing matrix (Maki-Nakagawa-Sakata matrix) 
$U_{MNS}$ shows very good agreement with the tri-bi maximal form \cite{tribi} implies that flavor structure 
is originated from a symmetry. 
Among them, non-Abelian discrete symmetries are well discussed as plausible possibilities \cite{review}.   

On the other hand, it has been established that about 23 $\%$ of energy density of the universe consists of Dark Matter (DM) \cite{Komatsu:2010fb}. Indirect detection experiments of DM, 
PAMELA \cite{Adriani:2008zr} and Fermi-LAT \cite{Abdo:2009zk,collaboration:2010ij}, 
reported excess of positron and the total flux $(e^++ e^-)$ in the cosmic ray. These observations can be explained by scattering and/or decay of 
TeV-scale DM particles. Since PAMELA measured negative results for anti-proton excess \cite{Adriani:2008zq}, 
leptophilic DM is preferable.

Even so, if the main final state of scattering or decay of DM is $\tau^+ \tau^-$,
 this annihilation/decay mode is disfavored because it will overproduce gamma-rays as final state radiation \cite{Papucci:2009gd}.
This may indicate that if the cosmic-ray anomalies are induced by DM scattering or decay, 
these processes also reflect flavor structure of the theory. 
There are several papers in which the DM nature is related to flavor symmetry \cite{a4-cosmic,Daikoku:2010ew,Hirsch:2010ru,Meloni:2010sk,Esteves:2010sh,Kajiyama:2006ww,Schmaltz:1994ws,Zhang:2009dd}.

While there are many models for the DM, we 
consider decaying DM model in this paper \cite{decay-compre,decay-gut,decay-others}. 
In such scenarios, no excess of anti-proton in the cosmic ray \cite{Adriani:2008zq} implies that 
lifetime of the DM particle should be of ${\cal O}(10^{26})$ sec. This long lifetime is achieved if 
the TeV-scale DM (gauge singlet fermion $X$) decays into leptons by dimension six operators 
$\bar L E \bar L X/\Lambda^2$ suppressed by GUT scale $\Lambda\sim 10^{16}~\gev$ \cite{decay-gut}. 
In this case, the lifetime of the DM is estimated as $\Gamma^{-1} \sim ((\tev)^5/\Lambda^4)^{-1}\sim 10^{26}$ sec. 
However, in general, there are gauge invariant dimension four decay operators which induce rapid DM decay,  and several dimension six operators which induce DM decay into quarks, Higgs and gauge bosons. 
Therefore one has to solve at least two problems for decaying DM models: i) why the lifetime of the DM is so long? 
ii) why the DM decays mainly into leptons? 
In ref. \cite{a4-cosmic}, we and our collaborators have shown that $A_4$ flavor symmetry can allow only 
the $\bar L E \bar L X/\Lambda^2$ operator and forbid the other undesirable operators, and determine flavor structure of DM decay mode. In that model, the $A_4$ symmetry allows only flavor universal leptonic decay mode, and the cosmic-ray anomalies are explained by fermionic DM decay. 

In this paper, we show that similar argument is possible in $T_{13}$ flavor symmetry model as 
an extension of the $A_4$ model. The $T_{13}$ group, isomorphic to $Z_{13} \rtimes Z_3$ group, is 
non-Abelian discrete subgroup of $SU(3)$. 
For the lepton sector, tri-bi maximal form can be derived by embedding three generations into 
triplet representations of flavor symetries. 
In this point of view, 
$A_4$ is the minimal group which has a triplet, and the $T_{13}$ group has two complex triplets 
in the irreducible representations. However,
since multiplication rules of the $T_{13}$ group is very different from those of $A_4$, 
we have a new texture of mass matrices of the lepton sector. 
As a result, although the $A_4$ model requires $SU(2)_L$ triplet Higgs bosons $\Delta$ with heavy mass $m_{\Delta}$ and small vacuum expectation values (VEVs) $v_{\Delta}$ to make the leptonic mixings and to suppress additional dimension five DM decay operator $H\Delta^{\dag}\bar L X$, 
our $T_{13}$ model does not require $\Delta$. The DM decay operators are also 
constrained by the $T_{13}$ symmetry so that only $\bar L E \bar L X/\Lambda^2$ operators are allowed like $A_4$ model of ref. \cite{a4-cosmic}. However unlike the $A_4$ model, DM decay mode depends on mixing matrices in general. In this paper we choose a particular set 
of parameters of the lepton sector, and show that the cosmic-ray anomalies can be well-explained by 
fermionic DM decay controlled by $T_{13}$ symmetry. 

This paper is organized as follows. We briefly discuss group theory 
of $T_{13}$ and lists the multiplication rules in the next section. 
In the section 3, we construct mass matrices of the lepton sector in definite choice of $T_{13}$ assignment of the fields, and show that there exists a consistent set of parameters. 
In the section 4, we show that only desirable dimension six DM decay operators are allowed by $T_{13}$ 
symmetry and that leptonic decay of the DM by those operators shows good agreement with 
the cosmic-ray anomaly experiments. The section 5 is devoted to the conclusions.

\section{$T_{13}$ group theory}
\label{sec:T7}
First of all, we briefly review the non-Abelian discrete group $T_{13}$, which is 
isomorphic to $Z_{13}\rtimes Z_3$ \cite{Fairbairn:1982jx,King:2009ap}.
The $T_{13}$ group is a subgroup of $SU(3)$, and known as the minimal
non-Abelian discrete group having two complex triplets as the irreducible representations.
We denote the generators of $Z_{13}$ and $Z_3$ by $a$ and $b$, respectively. 
They satisfy 
\begin{equation}
a^{13}=1,\quad ab=ba^9.
\end{equation}
Using them, all of $T_{13}$ elements are written as
\begin{equation}
g=b^{m}a^{n} ,
\end{equation}
with $m=0,1,2$ and $n=0,\cdots,12$.

The generators, $a$ and $b$, are represented e.g. as 
\begin{equation}
b=\Mat3{0}{1}{0} {0}{0}{1} {1}{0}{0},\quad 
a=\Mat3{\rho}{0}{0} {0}{\rho^3}{0} {0}{0}{\rho^9},
\end{equation}
where $\rho=e^{2i\pi/13}$.
These elements are classified into seven conjugacy classes,
\begin{eqnarray}
\begin{array}{ccc}
 C_1:&\{ e \}, &  h=1,\\
 C_{13}^{(1)}:&\{b~,~ ba~,~ba^{2}~,\quad...\quad,~ ba^{10}~,~ba^{11}~,~ba^{12}\}, &  
h=3,\\
  C_{13}^{(2)}:&\{b^2~,~ b^2a~,~b^2a^{2}~,\quad...\quad,~ b^2a^{10}~,~b^2a^{11}~,~b^2a^{12}\}, &  h=3,\\
 C_{3_1}:&\{ a~,~a^{3}~,~a^{9} \},&    h=13,\\
 C_{\bar3_1}:&\{ a^{4}~,~a^{10}~,~a^{12} \},&  h=13,\\
 C_{3_2}:&\{ a^{2}~,~a^{5}~,~a^{6} \},&    h=13,\\
 C_{\bar3_2}:&\{ a^{7}~,~a^{8}~,~a^{11} \},&  h=13.\\
\end{array}
\end{eqnarray}

The $T_{13}$ group has three singlets ${\bf 1}_k$ with $k=0,~1,~2$ and two complex triplets ${\bf 3_1}$ and ${\bf 3_2}$ as irreducible representations.
The characters are shown in Table \ref{T13}, where $\xi_1\equiv \rho+\rho^3+\rho^9$,  $\xi_2\equiv \rho^2+\rho^5+\rho^6$, and $\omega \equiv e^{2i\pi/3}$.

\begin{table}[t]
\begin{center}
\begin{tabular}{|c|c|c|c|c|c|c|c|c|c|}
\hline
        & $n$ & 
$h$&$\chi_{\bf 1_{0}}$&$\chi_{\bf 1_{1}}$&$\chi_{\bf1_{2}}$&$\chi_{\bf3_1}$&$\chi_{\bf\bar3_1}$&$\chi_{\bf3_2}$&$\chi_{\bf\bar3_2}$ \\ \hline
$C^{(0)}_1$  & $1$ &$1$&   $1$  &    $1$    &    $1$     &$3$ & $3$  &$3$ & $3$   \\ 
\hline
$C^{(1)}_{13}$  & $13$ &$3$&   $1$  &    $\omega$    &    $\omega^2$     
&$0$&$0$ &$0$&$0$  \\ \hline
$C^{(2)}_{13}$ & $13$  &$3$&   $1$  & $\omega^2$  & $\omega$ &$0$ & $0$  &$0$ & $0$   \\ 
\hline
$C_{3_1}$ &  $3$  &$13$&   $1$  & $1$  & $1$ &$\xi_1$ &$\bar\xi_1$ &$\xi_2$ &$\bar\xi_2$  \\ \hline
$C_{\bar3_1}$ & $3$ &$13$&   $1$  & $1$&  $1$  &$\bar\xi_1$ &$\xi_1$  &$\bar\xi_2$ &$\xi_2$  \\
\hline
$C_{3_2}$ &  $3$  &$13$&   $1$  & $1$  & $1$ &$\xi_2$ &$\bar\xi_2$ &$\xi_1$ &$\bar\xi_1$  \\ \hline
$C_{\bar3_2}$ & $3$ &$13$&   $1$  & $1$&  $1$  &$\bar\xi_2$ &$\xi_2$  &$\bar\xi_1$ &$\xi_1$  \\
\hline
\end{tabular}
\end{center}
\caption{Characters of $T_{13}$. $\bar\xi_i$ is defined as the complex conjugate of $\xi_i$.}
\label{T13}
\end{table}

Next we show the multiplication rules of the $T_{13}$ group. We define the triplets as
\begin{eqnarray}
{\bf3_1}\equiv\3tvec{x_1}{x_3}{x_9},\quad
{\bf\bar3_1}\equiv\3tvec{ \bar x_{12}}{ \bar x_{10}}{ \bar x_{4}},\quad
{\bf 3_2}=\3tvec{y_2}{y_6}{y_5},\quad
{\bf\bar3_2}\equiv\3tvec{ \bar y_{11}}{ \bar y_{7}}{ \bar y_{8}},
\end{eqnarray}
where the subscripts denote $Z_{13}$ charge of each element.

The tensor products between triplets are obtained as 
\begin{eqnarray}
\3tvec{x_1}{x_3}{x_9}_{{\bf 3_1}}\otimes\3tvec{y_1}{y_3}{y_9}_{{\bf 3_1}}
&=&
\3tvec{x_3y_{9}}{x_9y_{1}}{x_1y_{3}}_{{\bf \bar3_1}}\oplus
\3tvec{x_9y_{3}}{x_1y_{9}}{x_3y_{1}}_{{\bf \bar3_1}}\oplus
\3tvec{x_1y_{1}}{x_3y_{3}}{x_9y_{9}}_{{\bf 3_2}},
\\
\3tvec{\bar x_{12}}{\bar x_{10}}{\bar x_4}_{{\bf \bar3_1}}\otimes\3tvec{\bar y_{12}}{\bar y_{10}}{\bar y_4}_{{\bf \bar3_1}}
&=&
\3tvec{\bar x_{10}\bar y_{4}}{\bar x_4\bar y_{12}}{\bar x_{12}\bar y_{10}}_{{\bf 3_1}}\oplus
\3tvec{\bar x_4\bar y_{10}}{\bar x_{12}\bar y_{4}}{\bar x_{10}\bar y_{12}}_{{\bf 3_1}}\oplus
\3tvec{\bar x_{12}\bar y_{12}}{\bar x_{10}\bar y_{10}}{\bar x_4\bar y_{4}}_{{\bf \bar3_2}},
\\
\3tvec{x_1}{x_3}{x_9}_{{\bf 3_1}}\otimes\3tvec{\bar y_{12}}{\bar y_{10}}{\bar y_4}_{{\bf \bar3_1}}
&=&
\sum_{k=0,1,2} (x_1\bar y_{12}+\omega^k x_3\bar y_{10}+\omega^{2k} x_9\bar y_4)_{{\bf1}_k}
 \oplus
\3tvec{x_3\bar y_{12}}{x_9\bar y_{10}}{x_1\bar y_{4}}_{{\bf 3_2}}\oplus
\3tvec{x_1\bar y_{10}}{x_3\bar y_{4}}{x_9\bar y_{12}}_{{\bf \bar3_2}},\nn\\
\end{eqnarray}
\begin{eqnarray}
\3tvec{x_2}{x_6}{x_5}_{{\bf 3_2}}\otimes\3tvec{y_2}{y_6}{y_5}_{{\bf 3_2}}
&=&
\3tvec{x_5y_{6}}{x_2y_{5}}{x_6y_{2}}_{{\bf \bar3_2}}\oplus
\3tvec{x_6y_{5}}{x_5y_{2}}{x_2y_{6}}_{{\bf \bar3_2}}\oplus
\3tvec{x_6y_{6}}{x_5y_{5}}{x_2y_{2}}_{{\bf \bar3_1}},
\\
\3tvec{\bar x_{11}}{\bar x_{7}}{\bar x_8}_{{\bf \bar3_2}}\otimes\3tvec{\bar y_{11}}{\bar y_{7}}{\bar y_8}_{{\bf \bar3_2}}
&=&
\3tvec{\bar x_{8}\bar y_{7}}{\bar x_{11}\bar y_{8}}{\bar x_{7}\bar y_{11}}_{{\bf 3_2}}\oplus
\3tvec{\bar x_7\bar y_{8}}{\bar x_{8}\bar y_{11}}{\bar x_{11}\bar y_{7}}_{{\bf 3_2}}\oplus
\3tvec{\bar x_{7}\bar y_{7}}{\bar x_{8}\bar y_{8}}{\bar x_{11}\bar y_{11}}_{{\bf 3_1}},
\\
\3tvec{x_2}{x_6}{x_5}_{{\bf 3_2}}
\otimes
\3tvec{\bar y_{11}}{\bar y_{7}}{\bar y_8}_{{\bf \bar3_2}}
&=&
\sum_{k=0,1,2} (x_2\bar y_{11}+\omega^k x_6\bar y_{7}+\omega^{2k} x_5\bar y_8)_{{\bf1}_k}
 \oplus
\3tvec{x_6\bar y_{8}}{x_5\bar y_{11}}{x_2\bar y_{7}}_{{\bf 3_1}}\oplus
\3tvec{x_5\bar y_{7}}{x_2\bar y_{8}}{x_6\bar y_{11}}_{{\bf \bar3_1}},\nn\\
\end{eqnarray}
\begin{eqnarray}
\3tvec{x_1}{x_3}{x_9}_{{\bf 3_1}}\otimes\3tvec{y_2}{y_6}{y_5}_{{\bf 3_2}}
&=&
\3tvec{x_9y_{6}}{x_1y_{5}}{x_3y_{2}}_{{\bf 3_2}}\oplus
\3tvec{x_9y_{2}}{x_1y_{6}}{x_3y_{5}}_{{\bf \bar3_2}}\oplus
\3tvec{x_9y_{5}}{x_1y_{2}}{x_3y_{6}}_{{\bf 3_1}},\\
\3tvec{x_1}{x_3}{x_9}_{{\bf 3_1}}\otimes\3tvec{\bar y_{11}}{\bar y_{7}}{\bar y_8}_{{\bf \bar3_2}}
&=&
\3tvec{x_1\bar y_{11}}{x_3\bar y_{7}}{x_9\bar y_{8}}_{{\bf \bar3_1}}\oplus
\3tvec{x_3\bar y_{8}}{x_9\bar y_{11}}{x_1\bar y_{7}}_{{\bf \bar3_2}}\oplus
\3tvec{x_3\bar y_{11}}{x_9\bar y_{7}}{x_1\bar y_{8}}_{{\bf 3_1}},\\
\3tvec{x_2}{x_6}{x_5}_{{\bf 3_2}}\otimes\3tvec{\bar y_{12}}{\bar y_{10}}{\bar y_4}_{{\bf \bar3_1}}
&=&
\3tvec{x_2\bar y_{12}}{x_6\bar y_{10}}{x_5\bar y_{4}}_{{\bf 3_1}}\oplus
\3tvec{x_2\bar y_{10}}{x_6\bar y_{4}}{x_5\bar y_{12}}_{{\bf \bar3_1}}\oplus
\3tvec{x_5\bar y_{10}}{x_2\bar y_{4}}{x_6\bar y_{12}}_{{\bf 3_2}},\\
\3tvec{\bar x_{12}}{\bar x_{10}}{\bar x_4}_{{\bf \bar3_1}}\otimes\3tvec{\bar y_{11}}{\bar y_{7}}{\bar y_8}_{{\bf \bar3_2}}
&=&
\3tvec{\bar x_{4}\bar y_{8}}{\bar x_{12}\bar y_{11}}{\bar x_{10}\bar y_{7}}_{{\bf \bar3_1}}\oplus
\3tvec{\bar x_4\bar y_{7}}{\bar x_{12}\bar y_{8}}{\bar x_{10}\bar y_{11}}_{{\bf \bar3_2}}\oplus
\3tvec{\bar x_{4}\bar y_{11}}{\bar x_{12}\bar y_{7}}{\bar x_{10}\bar y_{8}}_{{\bf 3_2}}
.
\end{eqnarray}

The tensor products between singlets are obtained as 
\begin{eqnarray}
& & (x)_{{\bf1}_0}(y)_{{\bf1}_0}=(x)_{{\bf1}_1}(y)_{{\bf1}_2}=(x)_{{\bf1}_2}
(y)_{{\bf1}_1}=(xy)_{{\bf1}_0},~ \nonumber \\
& & (x)_{{\bf1}_1}(y)_{{\bf1}_1}=(xy)_{{\bf1}_2},~
(x)_{{\bf1}_2}(y)_{{\bf1}_2}=(xy)_{{\bf1}_1}.
\end{eqnarray}
\\
The tensor products between triplets and singlets are obtained as 
\begin{eqnarray}
(y)_{{\bf 1}_k}\otimes \3tvec{x(\bar x)_{1}}{x(\bar x)_{2}}{x(\bar x)_{3}}_{{\bf 3(\bar3)}}= 
\3tvec{yx(\bar x)_{1}}{yx(\bar x)_{2}}{yx(\bar x)_{3}}_{{\bf 3(\bar3)}} .
\end{eqnarray}

In the following section, we discuss mass matrices of the lepton sector 
determined by the $T_{13}$ flavor symmetry.

\section{Lepton masses and mixings}

In this section, we
discuss the lepton masses and mixings in the setup shown in 
Table~\ref{T13-assign}. Here, $Q$, $U$, $D$, $L$, $E$, $H (H')$ and $X$ denote left-handed quarks, 
right-handed up-type quarks, right-handed down-type quarks, left-handed leptons, 
right-handed charged leptons, Higgs bosons, and gauge singlet fermion, respectively. 
We introduce two $T_{13}$ triplet Higgs bosons $H({\bf 3_1})$ and 
$H({\bf \bar 3_2})$ which couple to leptons, and three $T_{13}$ singlet Higgs bosons 
$H' ({\bf 1_{0,1,2}})$ couple to quarks. This is realized by appropriate choice of 
an additional $Z_3$ symmetry. Since we concentrate on the lepton sector in this paper, 
$T_{13}$ charge of quarks and $H'$ are assigned to the singlets. Therefore, 
mass matrices in the quark sector are not constrained, while those 
in the lepton sector are determined by the $T_{13}$ symmetry.   
A gauge and $T_{13}$ singlet fermion $X$ is also 
introduced in addition to the SM matter fermions. Since the Yukawa couplings $LXH$ are 
forbidden by the $T_{13}$ symmetry, this new fermion $X$ does not work 
as right-handed neutrino, and left-handed Majorana neutrino mass terms are 
generated by dimension five operators $LHLH$. The fermion $X$ is 
a DM candidate decaying into leptons by dimension six operators $\bar LE \bar LX$. 
However, we postpone the discussion on this point to the next section.

For the matter content and the $T_{13}$ assignment given in
Table~\ref{T13-assign}, the charged-lepton and neutrino masses are generated 
from the $T_{13}$ invariant operators
\be
{\cal L}_{Y}&=&\sqrt{2}a_e \bar E L H^c ({\bf \bar 3_2})+
\sqrt{2}b_e \bar E L H^c ({\bf 3_1})\nn\\
&+&\frac{a_{\nu}}{\Lambda}LH({\bf \bar 3_2})LH({\bf \bar 3_2})
+\frac{b_{\nu}}{\Lambda}LH({\bf \bar 3_2})LH({\bf 3_1})
+\frac{c_{\nu}}{\Lambda}LH({\bf \bar 3_2})LH({\bf 3_1})+h.c.,
\label{lag}
\ee
where $H^c=\epsilon H^*$ and the fundamental scale $\Lambda=10^{11}~\gev$. 
After the electroweak symmetry breaking, the Lagrangian Eq.({\ref{lag}}) gives rise to mass matrices 
of charged leptons $M_e$ and neutrinos $M_{\nu}$ 
\be
M_e&=&\left( \begin{array}{ccc} 
0& b_e v_1 & a_e \bar v_2\\
a_e \bar v_3 &0& b_e v_2 \\
b_e v_3 &a_e \bar v_1&0 \\
\end{array}\right), ~
\label{me}\\
M_{\nu}&=&\frac{1}{\Lambda}\left( \begin{array}{ccc} 
c_{\nu}\bar v_3 v_2&a_{\nu} \bar v_1^2+b_{\nu}\bar v_3 v_1&a_{\nu}\bar v_3^2+b_{\nu}\bar v_2 v_3\\
a_{\nu} \bar v_1^2+b_{\nu}\bar v_3 v_1& c_{\nu}\bar v_1 v_3&a_{\nu}\bar v_2^2+b_{\nu}\bar v_1 v_2 \\
a_{\nu}\bar v_3^2+b_{\nu}\bar v_2 v_3&a_{\nu}\bar v_2^2+b_{\nu}\bar v_1 v_2&c_{\nu} \bar v_2 v_1\\
\end{array}\right), 
\label{mnu}
\ee
where the VEVs are defined as 
\be
\langle H({\bf 3_1})^i \rangle=\frac{v_i}{\sqrt{2}},~ 
\langle H({\bf \bar3_2})^i \rangle=\frac{\bar v_i}{\sqrt{2}},~
\sum_{i=1}^3 \left( v_i^2+\bar v_i^2\right)=(246~\gev)^2. 
\ee

\begin{table}[t]
\centering
\begin{tabular}{c|cccccccc} \hline\hline
& $Q$ & $U$ & $D$ & $L$ & $E$ & $H$ &$H'$& $X$ \\ \hline
$SU(2)_L \times U(1)_Y$ & 
{\bf 2}$_{1/6}$  & {\bf 1}$_{2/3}$ & {\bf 1}$_{-1/3}$ &
~{\bf 2}$_{-1/2}$~ & ~{\bf 1}$_{-1}$~  & {\bf 2}$_{1/2}$ &  ~{\bf 2}$_{1/2}$ &
~{\bf 1}$_0$~ \\
$T_{13}$ & 
${\bf 1_{0,1,2}}$ & ${\bf 1_{0,1,2}}$ & ${\bf1_{0,1,2}}$ &
${\bf 3_1}$ & ${\bf 3_2}$ &
${\bf 3_1},{\bf \bar 3_2}$ &${\bf1_{0,1,2}}$ &
${\bf 1_0}$ \\ 
$Z_3$ & 
$1$ & $\omega$ & $\omega^2$ &
$1$ & $1$ &
$1$ &$\omega$ &
$1$ \\ 
\hline
\end{tabular}
\caption{\small The $T_{13}$ and $Z_3$ charge assignment of the SM fields and the
dark matter $X$, where $\omega=e^{2i \pi/3}$. }
\label{T13-assign}
\end{table}

Now we give a numerical example. By the following choice of parameters
\be
v_1&=&0.118297~\gev,~v_2=179.257~\gev,~v_3=1.99413~\gev,\\ \nn
\bar v_1&=&10~\gev,~\bar v_2=168.164~\gev,~\bar v_3=0.0483626~\gev,\\ \nn
a_e&=&0.010566,~b_e=0,~a_{\nu}=-9.71512\times 10^{-5},~b_{\nu}=5.19702 \times 10^{-4},~
c_{\nu}=0.169389,
\ee
the mass matrices Eqs. (\ref{me}) and (\ref{mnu}) give rise to mass eigenvalues and 
related observatives as
\be
m_e&=&0.511~\mev,~m_{\mu}=105.66~\mev,~m_{\tau}=1776.82~\mev,\nn \\
m_{\nu 1}&=&1.385 \times 10^{-2}~\ev,~m_{\nu 2}=1.637 \times 10^{-2}~\ev,~
m_{\nu 3}=5.194 \times 10^{-2}~\ev, \\ \nn
\Delta m_{21}^2&=&m_{\nu 2}^2-m_{\nu 1}^2=7.59 \times 10^{-5}~\ev^2,~
\Delta m_{32}^2=m_{\nu 3}^2-m_{\nu 2}^2=2.43 \times 10^{-3}~\ev^2,~\\ \nn
\langle m\rangle_{ee}&=&1.47 \times 10^{-2}~\ev,~\sum_i m_{\nu i}=8.2 \times 10^{-2}~\ev,
\ee
and the mixing matrices
\be
U_{eL}&=&\left( \begin{array}{ccc}
1& 0&0\\
0&1&0\\
0&0&1\\
\end{array}\right),~
U_{eR}=\left( \begin{array}{ccc}
0& 0&1\\
1&0&0\\
0&1&0\\
\end{array}\right),~\label{uelr}\\ \nn
U_{MNS}&=&U_{eL}^{\dag}U_{\nu}=\left( \begin{array}{ccc}
0.828521 & 0.558869&0.0348995\\
-0.374962&0.60001&-0.706676\\
-0.41588&0.57241&0.706676\\
\end{array}\right),
\ee
which are all consistent with the present experimental data \cite{pdg}. In particular in the case of $U_{eL}=1$, 
the mass matrices Eqs. (\ref{me}) and  (\ref{mnu}) require 
normal hierarchy $m_{\nu 1}<m_{\nu 2}<m_{\nu 3}$ of the neutrino masses and $U_{e3}^{MNS}\neq 0$. 
A comprehensive analysis of the $T_{13}$ symmetry models will be 
published elsewhere \cite{homework}. 
In the following analysis, we assume these parameters to discuss decaying dark matter.

\section{Decaying dark matter in the $T_{13}$ model}

The cosmic-ray anomalies measured by PAMELA \cite{Adriani:2008zr} and Fermi-LAT \cite{Abdo:2009zk,collaboration:2010ij} can be explained by DM decay with lifetime of order 
$\Gamma^{-1} \sim 10^{26}$ sec. If the DM decays into leptons by dimension six operators $\bar L E \bar L X/\Lambda^2$, where $X$ is 
the gauge singlet fermionic DM, such long lifetime can be achieved.  
In general, however, there exist several gauge invariant decay operators of dimension four which induce 
rapid DM decay, and of dimension six which include quarks, Higgs and gauge bosons in the final states. In this section, we first show that the $T_{13}$ flavor symmetry forbids those undesired 
operators. Next we show by calculating the positron (electron) flux
that the scenario, which has a particular generation
structure of DM decay vertices, is possible to excellently describe
the cosmic-ray anomalies.

 \subsection{Dark matter decay operators}
 \bigskip
\begin{table}[h]
\centering
\begin{tabular}{ccl} \hline\hline
Dimensions && \multicolumn{1}{c}{DM decay operators} \\ \hline
4 && $\bar{L} H^c X$ \\ 
5 && ~~~$-$ \\
6 && 
$\bar{L}E\bar{L}X$,
~~$H^\dagger\!H\bar{L}H^cX$,
~~$(H^c)^tD_\mu H^c\bar{E}\gamma^\mu X$, \\
&&
$\bar{Q}D\bar{L}X$,
~~$\bar{U}Q\bar{L}X$,
~~$\bar{L}D\bar{Q}X$,
~~$\bar{U}\gamma_\mu D\bar{E}\gamma^\mu X$, \\
&& 
$D^\mu H^c D_\mu \bar{L} X$,
~~$D^\mu D_\mu H^c\bar{L}X$,  \\
&& 
$B_{\mu\nu}\bar{L}\sigma^{\mu\nu}H^cX$,
~~$W_{\mu\nu}^a\bar{L}\sigma^{\mu\nu}\tau^aH^cX$  \\ \hline
\end{tabular}
\medskip
\caption{\small The decay operators of the gauge-singlet fermionic
dark matter $X$ up to dimension six. 
$B_{\mu\nu}$, $W_{\mu\nu}^a$, and $D_\mu$ are the field strength
tensor of hypercharge gauge boson, weak gauge boson, and the
electroweak covariant derivative.\bigskip}
\label{op}
\end{table}

In our model, as mentioned before, a gauge-singlet 
fermion $X$ is introduced as the dark matter particle in addition to the SM fermions.
By assuming that the baryon number is preserved at least at perturbative
level, it turns out that there exist various gauge invariant
operators up to dimension six shown in Table \ref{op} ~\cite{six-dim} . From the table, 
one finds that the dark matter $X$ can in general 
decay into not only leptons but also quarks, Higgs, and gauge bosons
at similar rates by dimension six operators. Furthermore, a rapid decay of DM is induced
if the dimension four Yukawa operator $\bar{L}H^cX$ is allowed. One
may try to impose an Abelian (continuous or discrete) symmetry to
prohibit unwanted decay operators, but it is shown in ref. \cite{a4-cosmic} that it does not work but 
$A_4$ flavor symmetry does. However in the $A_4$ model discussed in ref. \cite{a4-cosmic}, 
$SU(2)_L$ triplet Higgs bosons $\Delta$ are introduced in order to give the mixings of the lepton sector, and 
these give rise to dimension five DM decay operators. In order to avoid rapid DM decay by the 
dimension five operators, small VEVs $\langle \Delta \rangle$ and large mass $m_{\Delta}$ are 
required. On the other hand, lepton masses and mixings are generated only by $SU(2)_L$ doublet Higgs bosons in the $T_{13}$ model as shown in the section 3.

Remarkably, by the field assignment of Table \ref{T13-assign}, all the decay operators listed in Table~\ref{op} 
except for $\bar{L}E\bar{L}X$ are forbidden due to this single symmetry. 
Consequently the DM mainly decays into three leptons. With the notation
$L_i=(\nu_{i},\ell_{i})=(U_{eL})_{i \alpha}(\nu_{\alpha},\ell_{\alpha})$ 
and $E_i=(U_{eR})_{i \beta}E_{\beta}$ $(i=1,2,3,~\alpha,\beta=e,\mu,\tau)$, 
the four-Fermi decay interaction is
explicitly written as 
\begin{eqnarray}
{\cal L}_{\rm decay}
&=& 
\frac{\lambda}{\Lambda^2}\,\sum_{i=1}^3(\bar{L}_iE_i)\bar{L}_iX 
\,+\text{h.c.} \nn\\
&=&\frac{\lambda}{\Lambda^2}\sum_{i=1}^3 \sum_{\alpha, \beta,\gamma=e,\mu,\tau}
\left( U_{eL}\right)_{i \alpha}^* \left( U_{eR}\right)_{i\beta}\left( U_{eL}\right)_{i \gamma}^* 
\label{lag2}\\ \nn
&&\left[ \left( \bar \nu_{\alpha}P_R E_{\beta}\right)\left( \bar \ell_{\gamma}P_R X\right)
-\left( \bar\ell_{\gamma}P_R E_{\beta}\right)\left( \bar \nu_{\alpha}P_R X\right)\right]
 +\text{h.c.}.
\end{eqnarray}
As seen from Eq. ({\ref{lag2}}), decay mode of the DM particle $X$ depends on 
the mixing matrices $U_{e(L,R)}$, which are given in Eq. ({\ref{uelr}}). 

\subsection{Positron production from dark matter decay}

Next, we consider the branching fraction of the DM decay
through the $T_{13}$ invariant Lagrangian Eq.(\ref{lag2}). Due to the
particular generation structure, the dark 
matter $X$ decays into several tri-leptons final state with the
mixing-dependent rate.
The decay width of DM per each flavor 
($\Gamma_{\alpha \beta \gamma}\equiv \Gamma(X \to  \nu_{\alpha}\ell^+_{\beta}\ell^-_{\gamma})$) 
turns out to be  
\be
\Gamma _{\alpha \beta \gamma}&=&
\frac{|\lambda|^2 m_X^5}{3072 \pi^3 \Lambda^4}
\left( U_{\alpha \beta \gamma}+U_{\alpha\gamma \beta}\right),
\label{decaywidth}
\ee
where $m_X$ is the DM mass, and 
\be
U_{\alpha \beta \gamma}=
\left| \sum_{i=1}^3\left( U_{eL}\right)_{i \alpha}^* \left( U_{eR}\right)_{i\beta}\left( U_{eL}\right)_{i \gamma}^*\right|^2.
\ee
Here we have omitted the masses of charged leptons in the final
states. 
The flavor dependent factor $U_{\alpha \beta \gamma}$ gives a factor three if one takes the sum 
of flavor indices $\alpha,\beta$ and $\gamma$. Therefore, the branching fraction of each decay mode is given by
\be
BR(X\to \nu_{\alpha}\ell^+_{\beta}\ell^-_{\gamma})
=\frac{1}{6}\left( U_{\alpha \beta \gamma}+U_{\alpha\gamma \beta}\right).
\ee
The DM mass $m_X$ and the total decay width 
$\Gamma=\sum_{\alpha,\beta,\gamma}\Gamma_{\alpha\beta\gamma}$ 
are chosen to be free parameters in the following analysis. 

Given the decay width and the branching fractions, the positron
(electron) production rate (per unit volume and unit time) at the 
position $\vec{x}$ of the halo associated with our galaxy is evaluated as
\begin{eqnarray}
  Q(E,\vec{x}) \,=\, n_X(\vec{x}) \,\Gamma
  \sum_f {\rm Br}(X\to f) \bigg[\frac{dN_{e^\pm}}{dE}\bigg]_f,
\label{source}
\end{eqnarray}
where $[dN_{e^\pm}/dE]_f$ is the energetic distribution of positrons
(electrons) from the decay of single DM with the final 
state `$f$'. We use the PYTHIA code~\cite{Sjostrand:2006za} to
evaluate the distribution $[dN_{e^\pm}/dE]_f$. The DM number 
density $n_X(\vec{x})$ is obtained by the profile $\rho(\vec{x})$, the
DM mass distribution in our galaxy, through the 
relation $\rho(\vec{x})=m_X n_X(\vec{x})$. In this work we adopt the
Navarro-Frank-White profile~\cite{Navarro:1996gj},
\begin{eqnarray}
  \rho_{\rm NFW}(\vec{x}) \,=\,
  \rho_\odot\frac{r_\odot(r_\odot+r_c)^2}{r(r+r_c)^2},
\end{eqnarray}
where $\rho_\odot\simeq0.30$~GeV/cm$^3$ is the local halo density
around the solar system, $r$ is the distance from the galactic center
whose special values $r_\odot\simeq8.5$~kpc and $r_c\simeq 20$~kpc
are the distance to the solar system and the core radius of the
profile, respectively.

As seen from Eqs. (\ref{uelr}) and (\ref{decaywidth}), the DM decays into 
$\tau^{\pm}$ as well as $e^{\pm}$ and $\mu^{\pm}$ in the equal rate. 
However, $\tau^{\pm}$ in the final states decay into hadrons, and such hadronic decays are suppressed by the electroweak coupling and the phase space factor.
As a result, pure leptonic decays give dominant contributions, and 
it is consistent with no anti-proton excess of the PAMELA results \cite{Adriani:2008zq}. 
On the other hand, the
injections of high-energy positrons (electrons) in the halo give rise
to gamma rays through the bremsstrahlung and inverse Compton
scattering processes. 
These gamma-ray flux~\cite{Abdo:2010nz} from leptonically decaying DM 
may constrain mass and lifetime of the DM~{\cite{Dugger:2010ys,Abazajian:2010zb}}. 
We will discuss this point later.
Therefore, we concentrate on calculating the $e^{\pm}$ fluxes in what follows. 
We follow ref. {\cite{a4-cosmic}} for diffusion model describing the propagation of positrons and electrons { \cite{Baltz:1998xv,Hooper:2004bq,Maurin:2001sj}}, and backgrounds {\cite{Baltz:1998xv,Pallis:2009ed}}.

\subsection{Results for PAMELA and Fermi-LAT}

\begin{figure}[t]
\begin{center}
\includegraphics[scale=0.55]{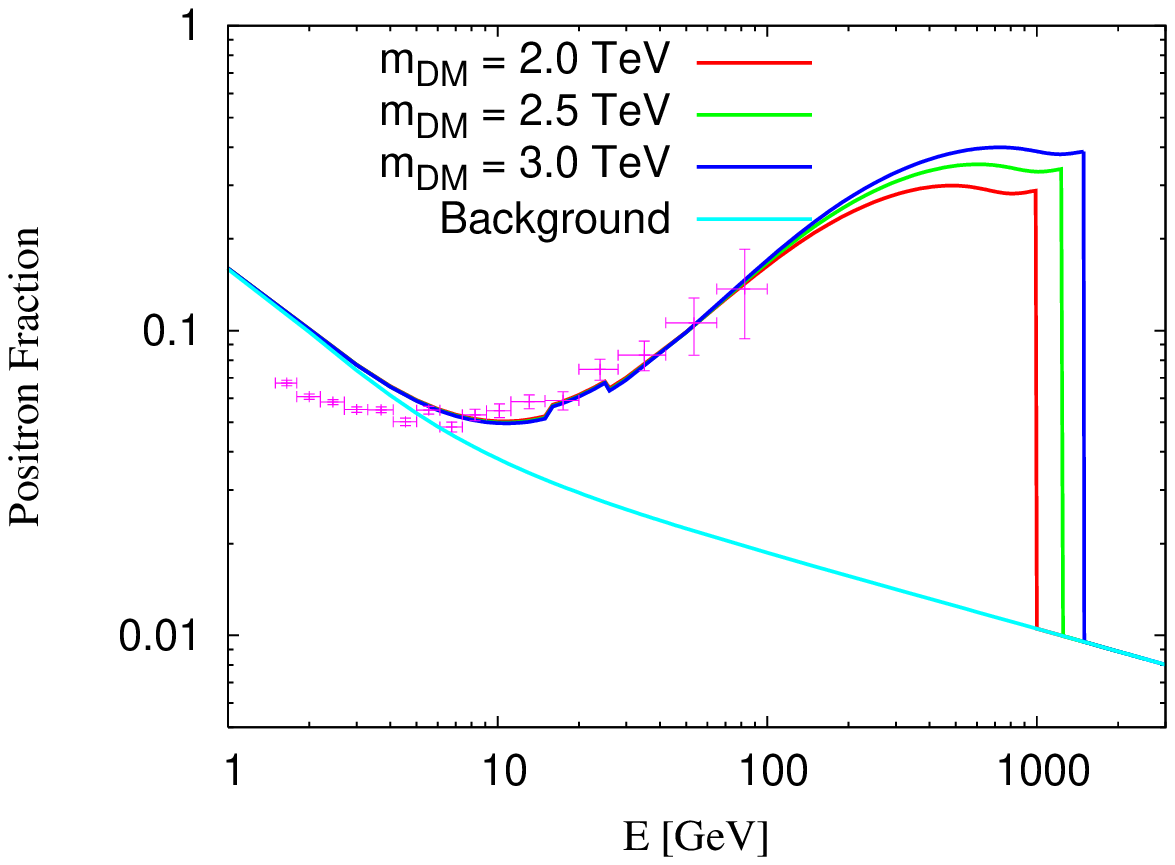}
\qquad
\includegraphics[scale=0.55]{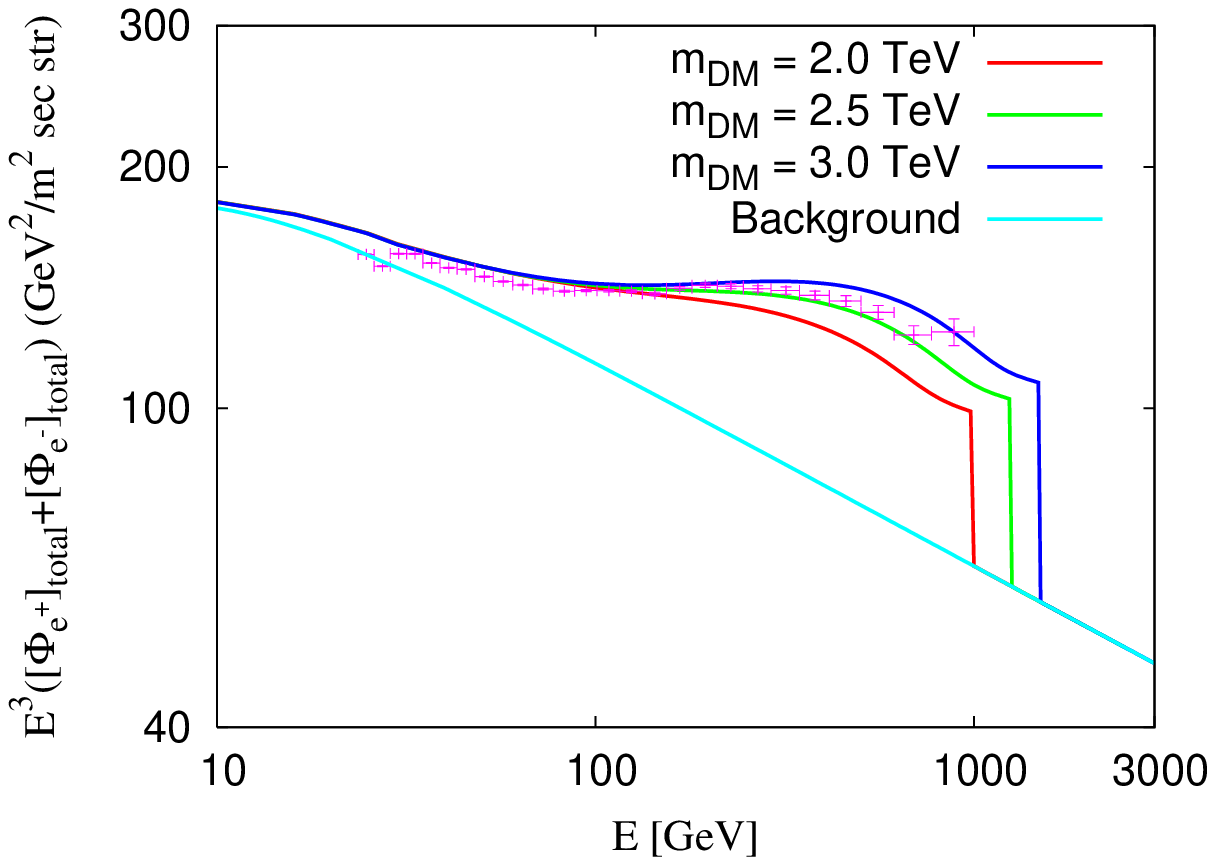}
\caption{\small The positron fraction \cite{Adriani:2008zr} and the total $e^++e^-$ flux \cite{Abdo:2009zk,collaboration:2010ij}
predicted in the leptonically-decaying DM scenario with 
$T_{13}$ symmetry. The DM mass is fixed to 2.0, 2.5, and 3.0~TeV\@. As for
the DM decay width used in the fit, see the text.}   
\label{fig:results}
\end{center}
\end{figure}

The positron fraction and the total 
flux $[\Phi_{e^-}]_\text{total}+[\Phi_{e^+}]_\text{total}$ are
depicted in Figure~\ref{fig:results} for the scenario of the
leptonically decaying DM with $T_{13}$ symmetry. 
For the DM mass $m_X=2$, $2.5$, and $3$~TeV, the results are shown
with the experimental data of PAMELA and Fermi-LAT\@. The total decay 
width $\Gamma$ is fixed for each value of DM mass so that the best fit
value explains the experimental data. With a simple $\chi^2$ analysis, we obtain
$\Gamma^{-1}=8.4\times10^{25}$, $6.95\times10^{25}$, and $5.9\times10^{25}$~sec 
for $m_X=2.0$, $2.5$, and $3.0$~TeV, respectively. 
One can see that in the $m_X=2.5$ TeV case, both experiments are well explained in  
the $T_{13}$ model.
Recent studies~{\cite{Dugger:2010ys,Abazajian:2010zb}} suggest that 
mass and lifetime of the decaying DM are 
strongly constrained by gamma-ray measurement
from cluster of galaxies, and that
allowed region which can simultaneously explain
the PAMELA and Fermi-LAT results does not exist.
To avoid these constraints, mass and lifetime
of the DM should be lighter than $\sim {\cal O}(\tev)$
and longer than $\sim {\cal O}(10^{27})$ sec, respectively.
In that case, only the PAMELA results can be explained by the decaying DM.

\section{Conclusions}

We have considered a new flavor symmetric model based on a non-Abelian discrete 
symmetry $T_{13}$. The $T_{13}$ group, isomorphic to $Z_{13}\rtimes Z_3$, is the minimal group which contains two complex triplets in the irreducible representations. 
The form of mass matrices are determined by the assignment of $T_{13}$ charges and  the multiplication rules.   We have shown that masses and mixings in the lepton sector are derived in the $T_{13}$ model consistently. 
Thanks to the complexities of the $T_{13}$ group compared to $A_4$, 
both the leptonic masses and mixings are made only by $SU(2)_L$ doublet Higgs bosons.  

We have also shown that the decay of gauge-singlet fermionic dark matter can explain the
cosmic-ray anomalies reported by the PAMELA and Fermi-LAT experiments. 
It is known that if the dark matter is TeV-scale fermionic particle, 
its longevity of order $10^{26}$ sec can be derived from dimension six four-fermi operators 
suppressed by a large scale of new physics. 
The $T_{13}$ symmetry forbids DM decay of final states with quarks, Higgs and gauge bosons, and 
allows only leptonic decay. Moreover, it determines the DM decay mode so that tauon final state does not give dominant contribution. We found that due to the fermionic DM decay controlled by the $T_{13}$ flavor symmetry, the cosmic-ray anomalies are well-explained.  

In this paper, we have explicitly given a numerical example of one consistent set of parameters in 
the mass matrices of the lepton sector. 
For completeness, a comprehensive analysis of mass matrices and its phenomenology in $T_{13}$ symmetric models will be published elsewhere \cite{homework}. 
\newpage
\vspace{0.5cm}
\hspace{0.2cm} {\bf Acknowledgments}
\vspace{0.5cm}

We would like to thank N. Haba, S. Matsumoto, M. Raidal and K. Yoshioka for 
useful discussions. 
The work of Y.K.\ was supported by the ESF grant No.~8090 and 
Young Researcher Overseas Visits Program for Vitalizing Brain Circulation  
Japanese in JSPS. H.O.\ acknowledges
partial supports from the Science and Technology Development Fund
(STDF) project ID 437 and the ICTP project ID 30.


\end{document}